\documentclass[tighten,twocolumn]{aastex631}
\usepackage{graphicx}
\usepackage{ulem}
\usepackage{amsmath}
\usepackage{wrapfig}
\usepackage[thinlines]{easytable}
\usepackage{tikz}
\usepackage{hyperref}
\usepackage{xcolor}
\usepackage{ulem}
\usepackage{booktabs}

\usepackage{float}
\begin{document}

\shortauthors{K{\i}ro\u{g}lu et al.}

\title{Black Hole Accretion and Spin-up through Stellar Collisions in Dense Star Clusters}

\correspondingauthor{Fulya K{\i}ro\u{g}lu}
\email{fulyakiroglu2024@u.northwestern.edu}

\author[0000-0003-4412-2176]{Fulya K{\i}ro\u{g}lu}
\affiliation{Center for Interdisciplinary Exploration \& Research in Astrophysics (CIERA) and Department of Physics \& Astronomy \\ Northwestern University, Evanston, IL 60208, USA}

\author[0000-0002-4086-3180]{Kyle Kremer}
\affiliation{Department of Astronomy \& Astrophysics, University of California, San Diego; La Jolla, CA 92093, USA}

\author[0000-0001-7616-7366]{Sylvia Biscoveanu}\thanks{NASA Einstein Fellow}
\affiliation{Center for Interdisciplinary Exploration \& Research in Astrophysics (CIERA) and Department of Physics \& Astronomy \\ Northwestern University, Evanston, IL 60208, USA}

\author[0000-0002-0933-6438]{Elena González Prieto}
\affiliation{Center for Interdisciplinary Exploration \& Research in Astrophysics (CIERA) and Department of Physics \& Astronomy \\ Northwestern University, Evanston, IL 60208, USA}

\author[0000-0002-7132-418X]{Frederic A. Rasio}
\affiliation{Center for Interdisciplinary Exploration \& Research in Astrophysics (CIERA) and Department of Physics \& Astronomy \\ Northwestern University, Evanston, IL 60208, USA}

\begin{abstract}
Dynamical interactions in dense star clusters could significantly influence the properties of black holes, leaving imprints on their gravitational-wave signatures. While previous studies have mostly focused on repeated black hole mergers for spin and mass growth, this work examines the impact of physical collisions and close encounters between black holes and (noncompact) stars.  Using Monte Carlo $N$-body models of dense star clusters, we find that a large fraction of black holes retained upon formation undergo collisions with stars. 
Within our explored cluster models, the proportion of binary black hole mergers affected by stellar collisions ranges from $10\%-60\%$. 
If all stellar-mass black holes are initially nonspinning, 
we find that up to $40\%$ of merging binary black holes may have components with dimensionless spin parameter $\chi\gtrsim 0.2$ because of prior stellar collisions, while typically about $10\%$ have spins near $\chi = 0.7$ from prior black hole mergers. We demonstrate that young star clusters are especially important environments as they can produce collisions of black holes with very massive stars, allowing for significant spin-up of the black holes through accretion. Our predictions for black hole spin distributions from these stellar collisions highlight their sensitivity to accretion efficiency, underscoring the need for detailed hydrodynamic calculations to better understand the accretion physics following these interactions.

\end{abstract}
 
\section{Introduction}

Gravitational-wave (GW) observations by the LIGO-Virgo-KAGRA (LVK) detectors have yielded vital insights into the properties of more than several hundred black holes (BHs), including their masses, spins, and distribution over redshift \citep{Abbott_2019,Abbott_2021,Abbott_2023}. Despite the growing number of BH binaries detected through GWs, their origins remain uncertain. 
Various mechanisms proposed for the formation of binary black holes (BBHs) include isolated evolution of massive binary stars in galactic fields, dynamical assembly in dense star clusters, migration and capture processes within active galactic nucleus (AGN) disks, and secular interactions in hierarchical triple systems \citep[see][for a review]{Mandel_2022}.
Each of these formation channels features its own set of poorly understood physical assumptions, presenting a major challenge for models attempting to produce robust predictions for the merger rates and properties of BHs. 
   
One key parameter that may help distinguish the contribution from these various formation channels is the BH spin \citep{Farr_2017,Farr_2018,Adamcewicz_2024}. The best-measured spin parameters using GW observations are the effective inspiral spin $\chi_{\rm eff}$ \citep{Ajith_2011} and the precessing spin $\chi_{\rm p}$ \citep{Wysocki_2019} defined as
\begin{align}
    \chi_{\rm eff} &= \frac{\chi_1 \cos{\theta_1} + \chi_2 \cos{\theta_2}}{(1+q)} \in (-1,1) \label{eq:chi_eff} \\
    \chi_{\rm p} &=  \max \left[ \chi_1 \sin{\theta_1}, \left(\frac{3+4q}{4+3q}\right)q \chi_2 \sin{\theta_2}\right]\in (0,1),
\end{align}
where $q\equiv m_2/m_1$ is the binary mass ratio, $m_1 \geq m_2$ are the BH masses, and $\chi_i\equiv S_i/m_i^2 \in (0,1)$ denotes the dimensionless spin of each BH, with $S_i $ representing the spin angular momentum of the BH. The effective inspiral spin  encodes a mass-weighted projection of the spin vectors on the orbital angular momentum axis, whereas the precessing spin reflects the projection of the spin vector on the plane of the orbit, capturing the spin-precessing effects.

Recent studies \citep[e.g.,][]{Rodriguez_2016,Rodriguez_2019} of dynamical assembly from star clusters have attempted to provide more robust predictions regarding BH spin magnitudes as these BHs are expected to exhibit isotropically distributed spins as a natural consequence of dynamical formation in gas-poor environments, assuming that most BHs are born slowly spinning from single star evolution. Yet, the spins of BHs at birth remain poorly understood \citep[e.g.,][]{Heger_2005,Lovegrove_2013,Qin_2019,Zevin_2022}. Theoretical studies suggest that single stellar-mass BHs typically have negligible birth spins ($\chi \approx 0.01$) due to efficient angular momentum transport from their progenitor stars' cores \citep{Fuller_2019}. Consistent with this, GW observations indicate that many of the BBH mergers detected by the LVK likely involve BHs with low spins \citep{Roulet_2019,Abbott_2023}. Nevertheless, spin can be acquired through stellar evolutionary processes such as mass transfer and tidal torques in close binaries \citep[e.g.,][]{Qin_2018,Ma_2019,Bavera_2020,Fuller_2022}, as well as through gas accretion \citep{Lopez_2019} and hierarchical mergers \citep{Antonini_2016,Rodriguez_2019} in dense star clusters. Similarly, the AGN disk channel can lead to spinning BHs through a combination of dynamics and gas interactions \citep{Vajpeyi_2022,McKernan_2024}. In fact, the latest LVK population analysis of GW events shows that the distribution of spin magnitudes among the most rapidly rotating components of BBHs detected in GWTC-3 peaks near $\chi \approx 0.4$ \citep[][see Figure 17]{Abbott_2023}. However, each formation channel provides different mechanisms for both spin-up and spin alignment of BBH binaries that yield contrasting predictions to compare with observations.

The latest catalog of GW detections (GWTC-3) reveals weak evidence for a preference against a purely isotropic spin distribution \citep[][see Figure 16]{Abbott_2023}, though the spin tilt distribution is only weakly constrained, and results are model dependent \citep{Vitale_2023}.
This could challenge the notion that all LVK BBHs originate from dynamical assembly in dense star clusters \citep{Rodriguez_2016,Rodriguez_2021,Stevenson_2017,Talbot_2017,Vitale_2017,Yu_2020}.

Additionally, isolated binary evolution predicts predominantly spin-orbit aligned BBH mergers due to past episodes of mass exchange and tidal interaction, which is also incompatible with observations. 
This has been interpreted as evidence for multiple formation channels \citep{Zevin_2021}. Alternatively, we will explore the possibility that additional physical processes that are not currently included in theoretical models may be necessary to explain the observed distribution of BH spins.

Globular clusters have been shown to efficiently produce merging BBHs through a series of close encounters in their very dense cores \citep[e.g.,][]{PortegiesZwart2000,Rodriguez_2016,Askar2017,Samsing_2018,Kremer2020_catalog,Mapelli_2021}.
The BH merger products of each of these sources inherit the angular momenta of their parent binaries. Assuming negligible component spins for ``first-generation'' (1G) BHs, these mergers produce a population of spinning BHs with $\chi \approx 0.7$ \citep{Berti_2008,Tichy_2008,Kesden_2010,Fishbach_2017}.
In some cases \citep[especially in more-massive star clusters with higher escape speeds;][] {Antonini_2016}, the merger product can be retained and merge again. 
The LVK event GW190521 has been touted as the canonical example of one of these ``second-generation'' (2G) BBH mergers.
In this case, GW events with high inferred spins may clearly indicate dynamical formation in dense star clusters \citep{Pretorius_2005,Fishbach_2017,Gerosa_2017,Rodriguez_2019,Doctor_2020,Kimball_2020,Gerosa_2021,Tagawa_2021,Payne2024}.

Additionally, BHs are expected to dynamically interact with luminous stars in star clusters. Indeed, the detection of a growing number of BH binaries with luminous stellar companions in both detached \citep{Giesers2018,Giesers2019} and accreting \citep[e.g.,][]{Strader2012,Chomiuk2013,Miller-Jones2015} configurations confirms the importance of such interactions. 
These encounters may cause a star to pass by a BH within its tidal disruption radius or to collide physically with the BH, resulting in the complete disruption of the star \citep[e.g.,][]{Perets_2016,Kremer2019_tde,Lopez_2019,Ryu_2020b,Kremer_2022,Kiroglu_2023}. Collisions often occur during close encounters involving binary stars (leading to three- or four-body resonant encounters) as shown in, e.g., \citet{Bacon1996,FregeauRasio2007,Kremer2021}. In cases where the disrupted stars are comparable in mass to the BH themselves, subsequent accretion could significantly increase the masses and spin magnitudes of the BHs involved. 
Encounters with such massive stars are expected to occur more frequently early in the evolution of clusters, while disruptions of low-mass stars will be more prevalent in older clusters where the most massive stars have already collapsed \citep[e.g.,][]{Kremer2019_tde}. 

Here, we explore how the initial properties of star clusters (metallicity, primordial binary fraction, and initial central density) affect the rates and types of close stellar interactions, ultimately influencing BH spin distributions.
Our paper is organized as follows. In Section \ref{sec:models} we summarize the methods used to model the long-term evolution of star clusters. In Section \ref{sec:colls} we discuss the numbers and properties of BH–star collisions identified in our models and discuss the dependence of these events on cluster properties. In Section \ref{sec:spin}, we detail our method for calculating BH spins from stellar collisions and subsequent accretion, presenting their distributions and discussing the effects of different cluster parameters on these spins. In Section \ref{sec:gw}, we examine the fraction of merging BBH populations influenced by these collisions, comparing the resulting spin distributions to the underlying astrophysical distributions inferred from LVK data. Finally, we conclude and discuss our results in Section \ref{sec:conc}.

\newpage

\begin{table}[ht]
\startlongtable
\begin{deluxetable*}{l|cc|cc|ccccc|ccc}
\label{table:IC}
\tablewidth{0pt}
\tablecaption{List of Cluster Models and Collision Counts\label{table:IC}}
\tablehead{
\multicolumn{3}{c}{} &
\multicolumn{2}{c}{\# BHs} &
\multicolumn{5}{c}{\# Collisions} &
\multicolumn{3}{c}{}\\
\cmidrule(lr){4-5}
\cmidrule(lr){6-10}
	\colhead{} &     \colhead{$^{1}r_v/\rm{pc}$} &   \colhead{$^{2}Z/Z_{\odot}$} &
\colhead{$^{3}$retained}&
\colhead{$^{4}$final}&
\colhead{$^{5}$MS}&
\colhead{$^{6}$Giant}&
\colhead{$^{7}$Merger prod.}&
\colhead{$^{8}$All}&
\colhead{$^{9}$Unique}&
\colhead{$^{10}\langle M_{\rm MS}\rangle$}&
\colhead{$^{11}\langle M_{\rm G}\rangle$}&
\colhead{$^{12}\langle M_{\rm BH}\rangle$}
}
\startdata
1 & 1.0 & 1.0 & 518 & 39 & 137 & 30 & 0 & 167 & 118 & 2.9 & 4.2 & 16.2 \\
2$^*$ & 1.0 & 1.0 & 253 & 2 & 168 & 73 & 0 & 241 & 120 & 2.5 & 3.8 & 15.2 \\
\hline
3 & 1.0 & 0.1 & 1754 & 573 & 46 & 24 & 5 & 70 & 67 & 0.5 & 23.6 & 29.0 \\
4$^*$ & 1.0 & 0.1 & 870 & 49 & 64 & 14 & 0 & 78 & 74 & 0.5 & 4.2 & 32.8 \\
\hline
5 & 0.5 & 1.0 & 698 & 13 & 366 & 134 & 4 & 500 & 265 & 2.3 & 4.9 & 14.7 \\
6$^*$ & 0.5 & 1.0 & 284 & 6 & 429 & 244 & 6 & 673 & 248 & 2.6 & 4.1 & 14.4 \\
\hline
7 & 0.5 & 0.1 & 1516 & 280 & 179 & 67 & 32 & 246 & 196 & 0.2 & 23.5 & 30.8 \\
8$^*$ & 0.5 & 0.1 & 834 & 3 & 189 & 58 & 6 & 247 & 177 & 1.3 & 9.8 & 23.5 \\
\enddata
\vspace{0.7cm}
\tablecomments
{All $N$-body models computed in this study. In Columns 1-2, we list the initial virial radius, metallicity. Models marked with asterisks assume $5\%$ for the binary fraction for massive stars ($>15\,M_{\odot}$), while other models assume $100\%$. We run each simulation for a minimum of $12\,{\rm Gyr}$. Columns 3-4 list the total number of BHs retained after natal kicks, and the number of BHs remained by the end of the simulation, respectively. Columns 5-6 list the total number of collisions of BHs with MS stars, and giants, respectively. Column 7 represents the number of BH-massive star collisions, where the massive star involved is itself a merger product of two progenitor stars, with at least one of the progenitors having a mass  $> 15\,M_{\odot}$. Column 8 represents the total number of BH--star collisions, encompassing all types of stars, while Column 9 counts only unique BHs involved in those collisions. Columns 10-12 list the median mass of MS stars, giant stars, and BHs (in units of $M_{\odot}$) that are involved in BH--star collisions. }
\end{deluxetable*}
\end{table}

\section{$N$-body Models of Clusters}
\label{sec:models}

We create eight cluster models in this study using the \texttt{Cluster Monte Carlo} (\texttt{CMC}) code, a H\'{e}non-style $N$-body code for stellar dynamics \citep[see][for a detailed review]{Rodriguez_2022}. \texttt{CMC} incorporates various physical processes essential for studying both formation and evolution of stellar-mass BHs, including stellar and binary star evolution using the \texttt{COSMIC} population synthesis package \citep{Breivik2020} which includes our most up-to-date understanding of the formation of compact objects, including prescriptions for natal kicks, mass-dependent fallback, and (pulsational) pair-instability supernovae \citep{Fryer_2001,Belczynski_2002}, three-body formation \citep{Morscher_2015}, and direct integration of small-$N$ resonant encounters \citep{FregeauRasio2007} including post-Newtonian effects \citep{Rodriguez_2018}.

All models we consider here assume $N=8\times10^5$ objects, including single and binary stars, at birth with masses drawn from an initial mass function (IMF) ranging from $0.08–150\,M_{\odot}$, following slopes of \cite{Kroupa2001}. Each model is initially described by King profiles \citep{King1962} with a fixed concentration parameter of $W_0=5$. We consider two metallicity values of $Z=[0.1, 1.0]\, Z_{\odot}$ and adopt a fixed galactocentric distance of 20 kpc in a Milky Way–like galactic potential. We also vary the initial cluster virial radius: $r_v=[0.5,1]\,\rm{\rm pc}$.

\begin{figure*}[ht!]
    \centering
    \includegraphics[width=0.9\linewidth]{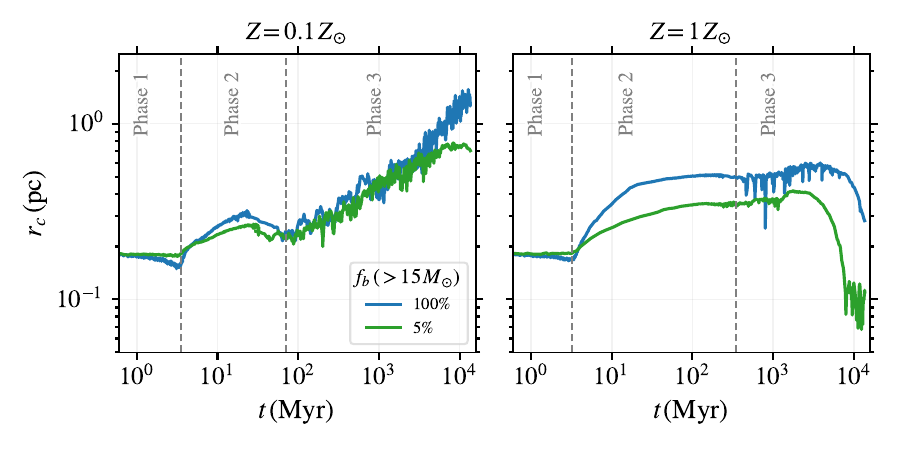}
    \caption{\footnotesize Time evolution of the cluster core radii \citep[measured as in][]{Chatterjee_2013} for models with metallicities $Z=0.1\,Z_{\odot}$ (left) and $Z=1\,Z_{\odot}$ (right), both with a fixed initial virial radius $r_v=1\,{\rm pc}$. In both panels, rapid mass segregation of massive MS stars is observed (labeled ``Phase 1''), with this effect being more pronounced and occurring more promptly at lower metallicity and higher binary fraction for massive stars. For all models, the core radius starts to expand around $t \approx 3\,{\rm Myr}$ due to mass loss from stellar evolution (``Phase 2''). On longer timescales ($t\gtrsim 100\,$Myr), the core expansion is further influenced by dynamical heating from BHs once a central BH core forms (``Phase 3''). In the case of $Z=1.0\,Z_{\odot}$, the rapid depletion of the BH population after 1~Gyr ultimately facilitates a significant core collapse, which in turn elevates the rate of BH--star collisions at later times (see the bottom panel of Figure \ref{fig:star_mass}).}  
        \label{fig:rc}
\end{figure*}
    \textit{Initial stellar binaries.} In dense star clusters, stellar binaries play an important role as a significant dynamical energy source, counteracting gravothermal contraction \citep[e.g.,][]{Heggie2003,Chatterjee_2010,Chatterjee_2013}. Binaries also significantly contribute to producing high rates of both stellar collisions \citep[e.g.,][]{FregeauRasio2007} and BH mergers \citep[e.g.,][]{Chatterjee2017a}. Observations indicate that nearly all O- and B-type stars in the Galactic field are born in binaries \citep{Sana_2012}. However, the binary fraction in star clusters—both primordial and at later stages—remains less well constrained. Many old globular clusters exhibit low binary fractions today, despite potentially higher primordial binary fractions at birth \citep[e.g.,][]{Fregeau_2009,Milone_2012}.
Motivated by this, we assume an initial low-mass ($<15\, M_{\odot}$) binary fraction of $5\%$ in all models. The initial binary fraction for massive stars ($15\,M_{\odot}$) stars is set to either $100\%$ or $5\%$. This variation is expected to alter the mass distribution of stars that could potentially collide with BHs, as models with $f_{\rm b}(>15\,M_{\odot})=100\%$ have been shown to produce very massive stars through runaway collisions \citep{Gonzalez_2021,Gonzalez_2024,Sharma_2024}.

For low-mass binaries, primary masses are drawn randomly from our IMF, secondary masses are drawn assuming a flat mass ratio distribution in the range [0.1, 1], and initial orbital periods are drawn from a log-uniform distribution $dn/ d \log P \propto P$.
For the secondaries of the massive stars $(>15 \,M_{\odot})$, a flat mass ratio distribution in the range [0.6, 1] is assumed, and initial orbital periods are drawn from the distribution $dn /d \log P \propto P^{-0.55}$ \citep[e.g.,][]{Sana_2012}. For all binaries, binary semi-major axes are drawn from near contact to the hard/soft boundary, and initial eccentricities are drawn from a thermal distribution.

\textit{Treatment of BH spins.} We assume that BHs are born with zero natal spin (dimensionless spin parameter $\chi=0$) motivated by the conclusions of \citet{Fuller_2019}. 
In our current version of \texttt{CMC}, BHs  acquire spin only through mergers with other BHs. In the event of a binary BH merger, we compute the spin (as well as mass and GW recoil kick) of the new BH using the method described in \cite{Rodriguez_2018}, which in turn implements phenomenological fits to numerical and analytic relativity calculations \citep{Campanelli_2007,Gonzalez_2007,Barausse_2009,Lousto_2013}.

\textit{Treatment of stellar collision products.} 
BHs in dense clusters can frequently undergo sufficiently close passages to tidally interact with stars. Depending on the pericenter distance ($r_p$), stars may undergo tidal captures, tidal disruptions, or physical collisions with BHs. Following \citet{Fabian1975}, we define the characteristic radius for tidal interaction as
\begin{equation}
\label{eq:rt}
    r_T = f_{\rm p}\,\left(\frac{M_{\rm BH}}{M_{\star}}\right)^{1/3}\,R_{\star},
\end{equation}
where $M_{\rm BH}$ is the BH mass, $M_{\star}$ and $R_{\star}$ are the stellar mass and radius, respectively, and $f_p$ is a dimensionless parameter that depends upon the internal structure of the star. \texttt{CMC} assumes that the star is instantaneously destroyed upon colliding with a BH, with no mass being accreted by the BH. This assumption arises from the expectation that most BHs interact with low-mass stars after the most-massive stars have already evolved in old globular clusters. However, in young star clusters, much more massive stars can be involved in collisions with BHs, and this assumption may no longer be justified.

\begin{figure}[ht]
    \centering
    \includegraphics[width=1\linewidth]{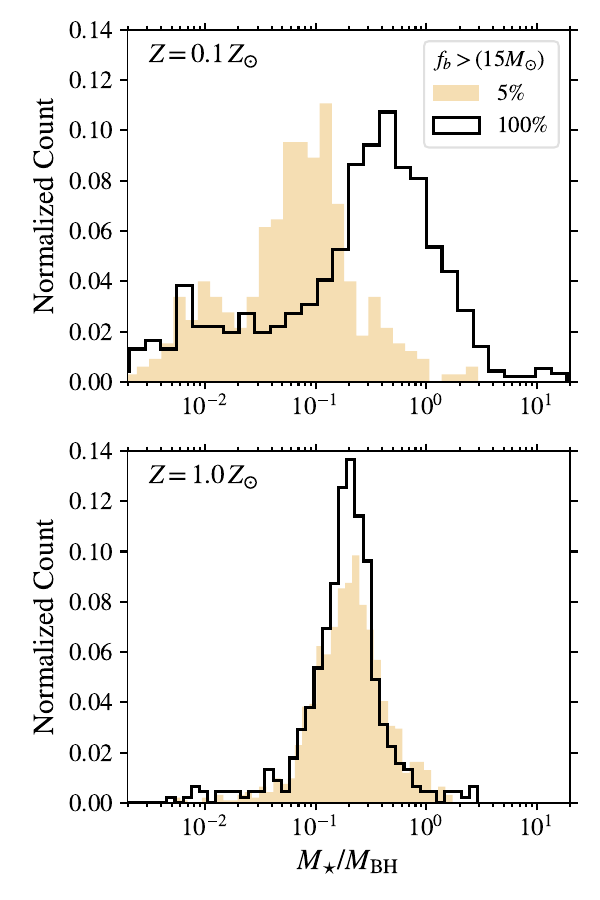}
    \caption{\footnotesize Mass ratio distribution for all BH--star collisions identified in models with high-mass binary fraction $5\%$ (beige histogram) vs. $100\%$ (black histogram) for different metallicity models. We see that BHs often collide with stars of comparable mass, especially for higher massive star binary fractions and lower metallicity.}
        \label{fig:mass_ratio}
\end{figure}

Predicting the outcomes of tidal interactions during close encounters requires detailed hydrodynamic calculations \citep[e.g.,][]{Kremer_2022,Ryu_2022,Kiroglu_2023}, which are beyond the computational scope of \texttt{CMC}. Consequently, we simplify the problem by assuming that stars are initially on parabolic orbits and are fully disrupted upon their first pericenter passage. 
As an upper limit, we set the pericenter distance to be $r_p=r_T$ computed by Equation~(\ref{eq:rt}) with $f_p=1$. We somewhat arbitrarily assume that each collision results in $50\%$ of star's mass remaining bound to the BH, with the actual accreted mass modified by an accretion efficiency factor \textbf{($0.5 \,M_{\star} \times f_{\rm acc}$)}, as discussed in Section \ref{sec:spin}.

\newpage 
\section{Demographics of Collisions from $N$-body Models}
\label{sec:colls}
\subsection{Rates and cluster evolution}
In this Section, we present the results of our cluster models. Table \ref{table:IC} lists the complete set of simulations with the initial conditions specified in the first two columns for each model. Columns 3 and 4 list the total number of BHs retained upon formation, and  by the end of the simulation, respectively. Column 8 shows the total number of BH--star direct collisions and tidal disruptions (which we simply refer to as ``collisions'' throughout the paper) that occur through both single–single and binary-mediated encounters involving main-sequence (MS; Column 5) and giant (Column 6) stars. Note that the total number of collisions can exceed the number of retained BHs upon formation, as some BHs (up to $10\%$) may experience multiple collisions with different stars over their lifetime in the cluster. To clarify, we also present the number of unique collisions in Column 9, which represents the count of distinct BHs colliding with stars. 

The collision rate can be simply estimated as  $\Gamma \approx n \Sigma \sigma_v N_{\rm BH}$, where  $n$ is the cluster number density, $\sigma_v$ is the cluster's velocity dispersion, $\Sigma$ is the collision cross section, and $N_{\rm BH}$ is the total number of BHs in the cluster \citep[which, in general, are all found within the half-mass radius due to mass segregation; e.g.,][]{Morscher_2015}.
For a given metallicity, the $r_v=0.5$ models yield a higher number of collisions (by at least a factor of 2) than those with $r_v=1$. This is expected, as the cluster density increases with decreasing cluster size, and the collision rate scales with the cluster number density $n \sim N/r_h^3$, where $r_h$ is the half-mass radius of the cluster.

Clusters with many retained BHs (e.g., models 3 and 7) are relatively diffuse while clusters with few BHs (models 2, 6) ultimately undergo core collapse leading to relatively high central densities and thus an increased rate of stellar collisions. For a given initial cluster size, the number of retained BHs upon formation in low-metallicity clusters is at least twice that of those in high-metallicity clusters. This difference stems from the stronger kicks assumed for lower-mass BHs in higher-metallicity models \citep[e.g.][]{Fryer_2012}, which eject about two-thirds of the BHs initially formed.

Nevertheless, despite having fewer retained BHs, high-metallicity clusters lead to BH--star collisions that are twice as frequent as those in lower-metallicity models.
This can be explained as follows: when a large number of BHs are present in a cluster, the energy production in the BH-dominated core dynamically ``heats'' the lower-mass stars in the cluster \citep[a phenomenon often referred to as ``BH binary burning'';][]{Mackey2007,BreenHeggie2013,Kremer2020_bhburning}. In this case, the cluster density is reduced, inhibiting interactions between BHs and stars \citep[e.g.,][]{Kremer_2018}. Thus, despite a significant number of retained BHs, the collision rate with stars remains relatively low. As the cluster depletes its BHs and approaches core collapse, the cluster density increases, and BHs dynamically mix more efficiently with other stars in the cluster \citep{Kremer2019}. 

The early dynamics of star clusters driven by massive stars is significantly influenced by the presence of primordial binaries \citep{Fregeau_2003}, the formation of binaries through three-body interactions \citep{Hut_1985,Giersz_1998}, and the mass-segregation process \citep[e.g.,][]{Freitag2006, Goswami2012}. In Figure~\ref{fig:rc}, we show how these various processes influence the evolution of cluster core radii and, in turn, the BH-star collision rate over time. Before the formation of BHs and their decoupling from the rest of the cluster, the most-massive stars tend to segregate toward the cluster center, which drives core collapse, which is more pronounced in models with $100\%$ massive binary fraction (Phase 1 in Figure \ref{fig:rc}).
The presence of binaries in the core further elevates collision rates, leading to successive collisions of massive stars within the first 5 Myr of cluster evolution. Shortly after formation of BHs, they can collide with these massive stars (see Column 7 of Table \ref{table:IC}). The number of massive stellar collisions is expected to be significantly higher in denser clusters with $f_{b}(>15\,M_{\odot})=100\%$ \citep[e.g.,][]{Gonzalez_2021}, resulting in an increased BH--massive star collision rate. Indeed, in model 7, we find that about $15\% $ of BH--star collisions involve the massive products of stellar collisions, potentially having a significant impact on BH properties, as discussed further in Section \ref{sec:spin}.

\subsection{Masses}

Columns 10-12 of Table~\ref{table:IC} present the median masses of BHs and stars (separately for MS stars and giants) involved in collisions. In the $Z=0.1\,Z_{\odot}$ clusters, more than $50\%$ of these collisions occur within $100\,\rm{Myr}$ with a similar median mass for both BHs and giants of $\approx 30\,M_{\odot}$. Approximately $20\%$ of the BHs exceed our assumed pair-instability supernova mass limit which is $40.5\,M_{\odot}$. About $25\% $ of these massive BHs result from BBH mergers earlier in the cluster's evolution (i.e., 2G BHs) while $75\%$ are formed through one or more stellar collisions of their progenitor stars in clusters with $f_{b}\,(>15\,M_{\odot})=100\%$ \citep[e.g.,][]{Kremer_2020,Gonzalez_2021}.

In contrast, the solar-metallicity models yield lower masses for both BHs and stars, with a typical median BH mass of $15\,M_{\odot}$. Higher metallicity leads to more substantial mass loss through stellar winds before the stellar core collapse, resulting in the formation of BHs with reduced masses \citep[e.g.,][]{Vink_2001,Fryer_2012,Belczynski_2016}.

\begin{figure}[t]
    \centering
    \includegraphics[width=1\linewidth]{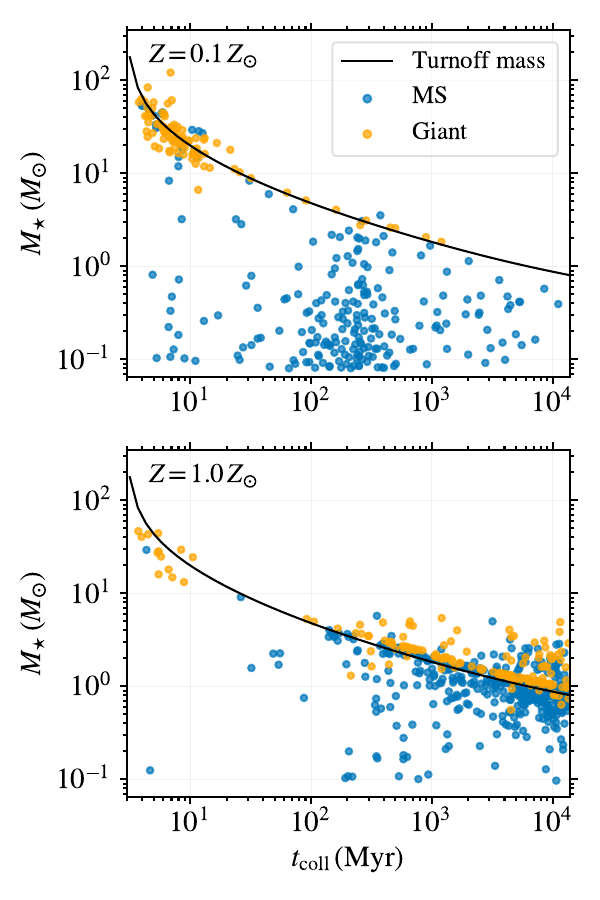}
    \caption{\footnotesize  Stellar mass vs. time of all BH-MS (blue) and BH-Giant (orange) collisions identified in models 1-8. Top (bottom) panel is for metallicity $0.1\,Z_{\odot} (1\,Z_{\odot})$. The solid black line shows the turnoff mass as a function of time (stars above this line were formed through previous stellar collisions/mergers). We see that BHs collide with massive stars (MS and giants) particularly at early times but also at late times for a cluster that undergoes core collapse (see the right panel of Figure \ref{fig:rc}).}
        \label{fig:star_mass}
\end{figure}

In Figure \ref{fig:mass_ratio}, we show the mass ratio distributions of colliding stars and BHs across models with different primordial binary fractions for massive stars ($f_{\rm b}(>15\,M_{\odot})=5\%,100\%$) and metallicities ($Z=0.1,1.0\,Z_{\odot}$). The top panel demonstrates that in clusters with $Z = 0.1\,Z_{\odot}$, increasing the primordial massive binary fraction shifts the median mass ratio to 0.5 and extends the tail for mass ratios to above 10. We note that the median BH mass remains relatively constant at approximately $30\,M_{\odot}$ in the low-metallicity models, regardless of the high-mass binary fraction (see Column 12 in Table \ref{table:IC}). Consequently, cases with $M_{\star}/M_{\rm BH}\gtrsim 1$ observed in $f_{\rm b}(>15\,M_{\odot})=100\%$ clusters are primarily due to massive stars formed through previous stellar mergers or collisions, which are mixed with BHs within the first roughly 10~Myr of cluster evolution (see the top panel of Figure \ref{fig:star_mass}). Additionaly, in models with $f_b (>15\,M_{\odot})=100\%$, $70\%$ of the massive star - BH collisions within 10~Myr involve binary encounters (binary-single or binary-binary), whereas this fraction decreases to $40\%$ for the models with $f_b (>15\,M_{\odot})=5\%$.
Most of the collisions ($90\%$) involving massive stars in the lower-metallicity models originate from primordial binaries, while only $10\%$ of these binaries are dynamically assembled through exchange encounters. However, we note that about $80\%$ of them are primordial binaries that have been dynamically shaped by a series of scatterings, during which their semi-major axis or eccentricity changes while retaining their original pairing. This highlights that the mass distribution of collided objects is highly sensitive to the primordial binary fraction of massive stars in low-metallicity models. Meanwhile, the bottom panel of Figure \ref{fig:mass_ratio} shows that in high-metallicity clusters, the mass ratio distribution of colliding objects remains largely unchanged despite variations in the massive binary fraction, with the distribution peaking around 0.2.

\begin{figure*}[t!]
    \centering
    \includegraphics[width=1\linewidth]{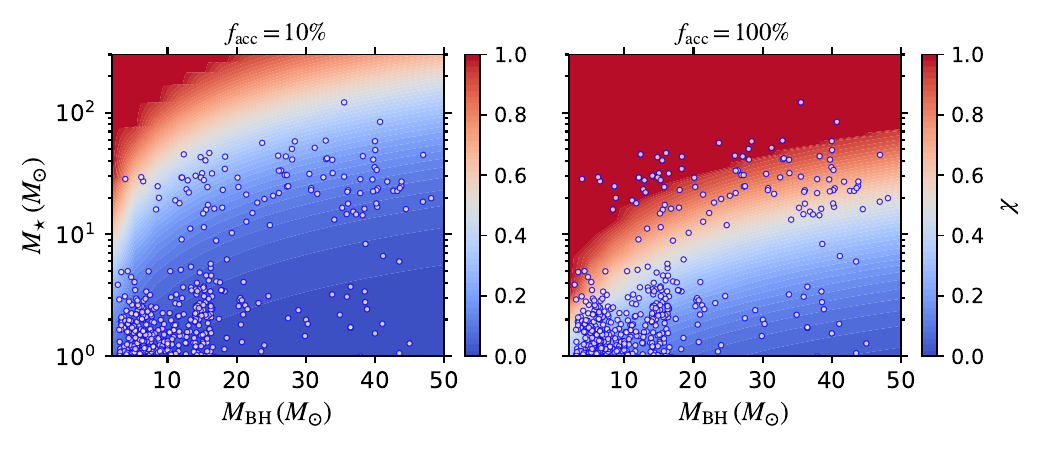}
    \caption{\footnotesize Dimensionless spin parameter $\chi$ for BHs of mass $M_{\rm BH}$ after a collision with star of mass $M_{\star}$. The dots show all collisions that happened in the models with $f_b (>15\,M_{\odot})=100\%$. In each panel, the background colors indicate the final spin of the BHs after accreting either $10\%$ (left) or $100\%$ (right) of the bound stellar debris, which we assume to be $0.5\,M_{\star}$. In the most optimistic scenario ($f_{\rm acc}=100\%$), stellar collisions can significantly affect BH spins, resulting in a median spin parameter  $\chi \approx 0.4$. For less-efficient accretion ($f_{\rm acc}=10\%$), the median BH spin is only about $\chi \approx 0.05$. Note again that we assume all BHs to be born initially nonspinning.}
        \label{fig:spin_contour}
\end{figure*}

In Figure \ref{fig:star_mass}, we show the stellar mass and collision time for all BH--star collisions in our models. In the lower-metallicity models (top panel), there is a distinct overdensity of collisions, especially of giant stars, shortly after formation of BHs, within $\lesssim 10\,\rm{Myr}$. In contrast, the solar-metallicity models (bottom panel) show a pronounced increase in collisions at $t\gtrsim 1$ Gyr. This late-time rise in collisions is attributed to the significant increase in cluster density as the cluster nears core collapse (see the right panel of Figure \ref{fig:rc}). In high-metallicity environments, massive giants experience significant mass loss due to stellar winds, leading to smaller radii and reduced collision cross sections. This reduction in cross section leads to fewer collisions early on. However, as lower-mass stars become more prevalent later, the impact of mass loss on collision rates diminishes. 
In fact, the increased density of the cluster during core collapse drives collisions between stars, leading to the formation of stars above the turnoff mass (indicated by the solid gray curve), such as blue straggler stars \citep[e.g.,][]{Bailyn1995}, which may subsequently collide with BHs.
This explains the constant mass ratio in star-BH collisions observed in Figure \ref{fig:mass_ratio} for high-metallicity models despite variations in the massive binary fraction, which primarily affect only the early collision dynamics within the cluster.

Our $N$-body models indicate that more than $50\%$ BHs retained upon formation in dense star clusters undergo at least one stellar collision throughout the cluster’s evolution. Notably, we found that approximately $35\%$ of these collisions have mass ratios greater than 0.5 in clusters with $f_{b}(>15\,M_{\odot}) = 100\%$ while this number decreases to a few percent in clusters with  $f_{b}(>15\,M_{\odot}) = 5\%$. This emphasizes that the presence of massive stars in binaries is crucial for achieving high rates of collisions between BHs and massive stars.

\section{Spins}
\label{sec:spin}

\subsection{Modeling Spin Evolution due to Accretion
}
We calculate the change of the BH spin magnitude through the accretion of disrupted material as described in \cite{Bardeen_1972}, where we assume that the final BH will have a mass and angular momentum nearly equal to those of the binary system at the last stable orbit (ISCO). Assuming that the accretion disk is in the equatorial plane of a BH and that gas is dumped directly onto the BH from the ISCO, the spin parameter of the BH is given by \cite{Bardeen_1972,Volonteri_2013}
\begin{equation}
\label{eq:chi}
    \chi = \begin{cases} 
    \frac{r^{1/2}}{3}q\left[4-\left(3 q^2 r-2\right)^{1/2}\right] & \rm{for}\quad \frac{1}{\it{q}} \leq \it{r}^{\rm{1/2}} \\
    1&  \rm{for}\quad \frac{1}{\it{q}} \geq \it{r}^{\rm{1/2}}
    \end{cases}
\end{equation}
where $q$ is the initial to final BH mass ratio and $r \equiv r_{\rm LSO}/M_{\rm BH}$ is the dimensionless radius of the ISCO in natural units, where $c=G=1$,
\begin{equation}
\label{eq:isco}
    r = 3 + Z_2 \mp \sqrt{(3-Z_1)(3+Z_1+2Z_2)},
\end{equation}
where $Z_1$ and $Z_2$ are functions of initial spin parameter $\chi_0$ only
\begin{equation}
      Z_1 \equiv 1+ (1-\chi_0^2)^{1/3}[(1+\chi_0)^{1/3}+(1-\chi_0)^{1/3}],\nonumber{}
\end{equation}
   \begin{equation}
   \label{eq:Z}
    Z_2  \equiv [3\chi_0^2+Z_1^2]^{1/2},   
\end{equation}
and the upper and lower signs indicate that the gas is accreting on prograde and retrograde equatorial orbits, respectively. 

In our cluster models, although BHs are initially born with zero spin, approximately 
$10\%$ of these BHs may experience \textit{more than one} stellar collision. As a result, some of these BHs could possess nonzero spin at the time of their second or subsequent collisions. Therefore, for our spin calculations, we use a generalized prescription that accounts for the possibility of an initially spinning BH. We however note that the collision of a BH with the most- massive star it has collided with is expected to have the most significant impact on the BH's spin.

According to Equation~(\ref{eq:chi}), an initially nonspinning BH ($\chi_0=0$) can be spun up to a maximum value $\chi = 1$ after its mass increases by a factor $\sqrt{6}\approx 2.5$. Conversely, a maximally rotating Kerr BH ($\chi_0=1$) can be spun down to $\chi = 0$ after growing by a factor $\sqrt{3}/2 \approx 1.2$. Spin-up occurs naturally when the captured material has a constant angular momentum axis aligned with the BH’s spin direction (“coherent accretion”). In contrast, spin-down happens when the accreted material has counterrotating angular momentum relative to the BH's spin. In our calculations, we assume that the accretion of corotating material (causing spin-up) and counterrotating material (causing spin-down) is equally probable. However, since the ISCO for retrograde orbits is at a larger radius than for prograde orbits, the transfer of angular momentum is however more efficient in the retrograde case. The accretion of counterrotating material is therefore more effective in spinning down BHs than the accretion of corotating material is in spinning them up.

\subsection{Spin Distributions under Realistic Stellar Collisions}
\label{sec:spin_calc}

We now investigate the spin evolution of BHs due to accretion, employing Equations (\ref{eq:chi}-\ref{eq:Z}) as a post-processing step to our $N$-body cluster models.

As shown in Equation~(\ref{eq:chi}), the final spin of a BH is determined by the amount of mass accreted, which depends on both the mass bound to the BH after the collision and the accretion efficiency, $f_{\rm acc}$. Motivated by the outcomes of our previous hydrodynamic simulations \citep{Kremer_2022,Kiroglu_2023}, we assume that the %accretion onto the BH is super-Eddington and the 
accretion flow is disk-like. For the highly super-Eddington accretion rates expected here, where the disk cannot efficiently cool via radiation, the disk is prone to outflows that reduce the total mass supplied to the BH \citep[e.g.,][]{NarayanYi1995,Blandford1999}. To account for potential mass outflows, the accretion rate onto the BH is adjusted by a factor of $(R_{\rm acc}/R_{d})^s$, where $R_{\rm acc} = 6GM_{\rm BH}/c^2$ represents the accretion radius (assumed to be the ISCO), $R_d$ is the outer edge of the disk, and the uncertain power-law index $s$ is in the range $[0,1]$. In the case of the highest mass inflow rate ($s=0$), the entire disk mass is accreted onto the BH ($f_{\rm acc} = 100\%$).
Conversely, for inefficient accretion disks ($s \approx 1$), most of the disk mass is expelled in a wind.

The true accretion efficiency depends on the complex physics of super-Eddington accretion, including magnetohydrodynamic processes and radiative transport, which are not fully understood \citep{Blaes_2011,Ohsuga_2011,McKinney_2014,Sadowski_2016}.
Numerical simulations of radiatively inefficient accretion flows analogous to those expected here suggest that $s$ most likely lies within the range $0.2$–$0.8$ \citep[e.g.,][]{Yuan2012,Yuan2014}. Considering an initial disk mass of $1\,M_{\odot}$, a disk radius of $R_d = 10^{11}\,\rm{cm}$, an accretion radius $R_{\rm acc} = 10^6\,\rm{cm}$, and high mass inflow rates ($s \lesssim 0.2$) leads to $(R_{\rm acc}/R_{d})^s\lesssim 0.1$.

Figure \ref{fig:spin_contour} illustrates the initial masses of BHs and stars involved in collisions, with the background color representing the corresponding BH spin for a given mass ratio, determined by the accretion efficiency. Given the uncertainties surrounding the accretion efficiency, we calculate upper and lower limits for BH spins, assuming accretion efficiencies of $10\%$ and $100\%$.

%\begin{table}[ht]
\startlongtable
\begin{deluxetable*}{l|cc|cccc|ccccc}
%\tabletypesize{\scriptsize}
\tablewidth{0pt}
\tablecaption{Merging Black Hole Binaries
\label{table:mergers}}
\tablehead{
    \multicolumn{3}{c}{} &
    \multicolumn{4}{c}{\# BBH Mergers} &
	\multicolumn{4}{c}{\# Spinning Components}\\
      \cmidrule(lr){4-7}
  \cmidrule(lr){8-11}
	\colhead{} &
     \colhead{$^{1}r_v/\rm{pc}$} &
    \colhead{$^{2}Z/Z_{\odot}$} &
    \colhead{$^{3}$No Colls} &
    \colhead{$^{4}$Single BH--star} &
    \colhead{$^{5}$BBH--star} &
    \colhead{$^{6}$Total} &
    \colhead{$^{7}$Primary} &
	\colhead{$^{8}$Secondary} &
 	\colhead{$^{9}$Both} & 	
    \colhead{$^{10}$Total}
	} 
\startdata
1 & 1.0 & 1.0  & 28 & 19 & 1 & 48 & 11 & 7 & 5 & 23 \\
2$^*$ & 1.0 & 1.0  & 8 & 13 & 6 & 27 & 9 & 2 & 10 &21 \\
\hline
3 & 1.0 & 0.1 & 87 & 10 & 0 & 97 & 9 & 1 & 3 &13\\
4$^*$ & 1.0 & 0.1  & 59 & 17 & 0 & 76 & 17 & 6 & 0&23 \\
\hline
5 & 0.5 & 1.0  & 37 & 48 & 5 & 90 & 26 & 9 & 21 & 56\\
6$^*$ & 0.5 & 1.0  & 6 & 30 & 13 & 49 & 16 & 4 & 25 &45 \\
\hline
7 & 0.5 & 0.1 & 58 & 29 & 0 & 87 & 24 & 5 & 9 & 38 \\
8$^*$ & 0.5 & 0.1  & 62 & 40 & 2 & 104 & 37 & 7 & 12  &56\\
\hline
\enddata
\vspace{0.2cm}
\tablecomments{In Columns 1-2, we list the initial cluster parameters—virial radius, metallicity, respectively—for the same set of models presented in Table \ref{table:IC}. Column 3 represents the number of mergers of BH binaries that are not influenced by stellar collisions, including both 1G and 2G components. Column 4 provides the number of mergers of BH binaries with at least one component that was spun up from stellar collisions prior to forming the BH binary and merger (Case B in Figure \ref{fig:cartoon}). Column 5 specifically represents the number of merging BH binaries that previously underwent a stellar collision as a binary and were compact enough to merge subsequently, ensuring that any temporary spin alignment was not erased before coalescence  (Case A in Figure \ref{fig:cartoon}). Column 6 lists the total number of BBH mergers, including those occurring within their host clusters and after their ejection, which is the sum of the values in Columns 3-5. Columns 7-9 represent the number of merging BH binaries with primary, secondary, or both components spinning, respectively, due to either repeating BBH mergers or stellar collisions. Column 10 represents the total number of these BBH mergers that have at least one component spinning. See Figure \ref{fig:spin_dist} for the primary spin distribution of the merging BH binaries.
}
\end{deluxetable*}

\begin{figure}[t!]
   \centering
\includegraphics[width=0.83\linewidth]{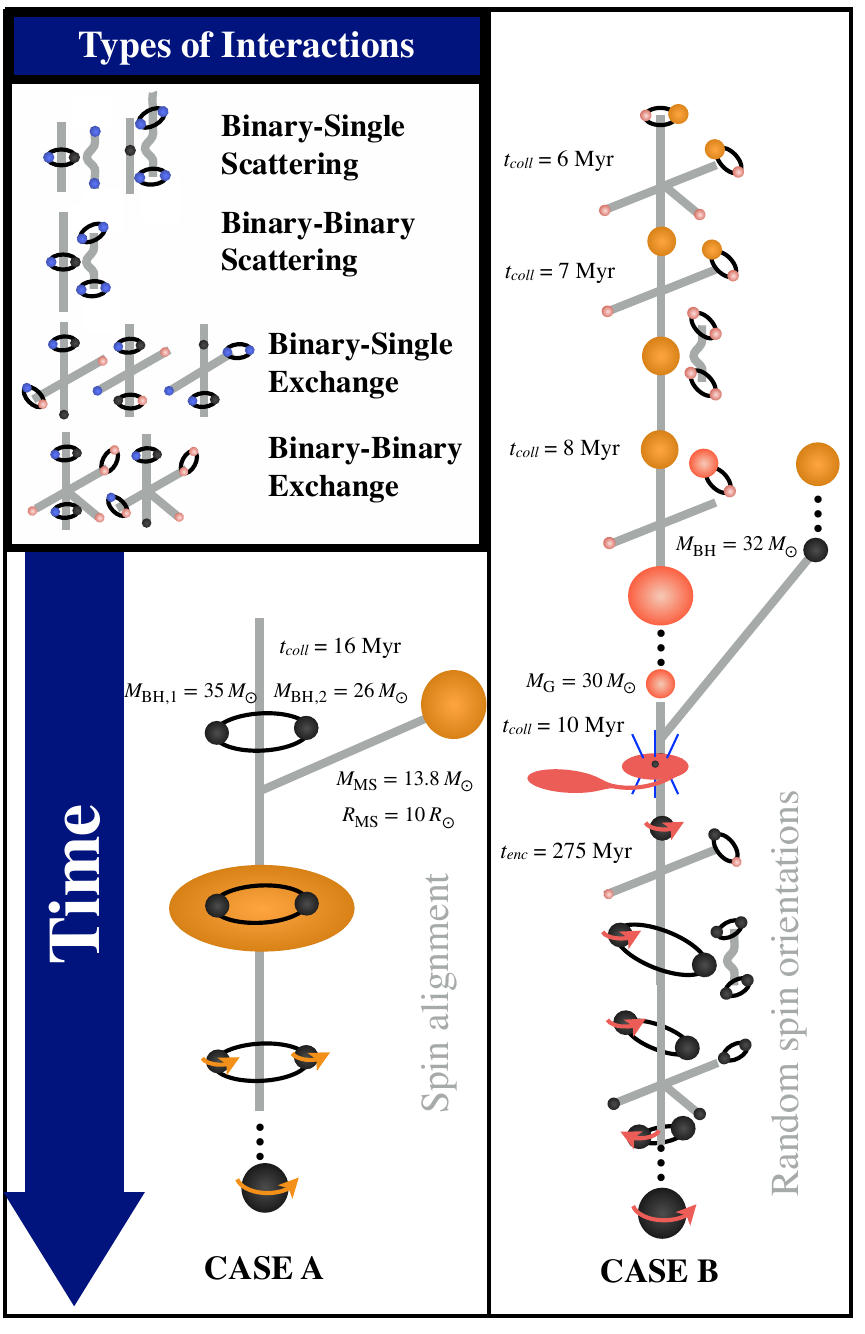}
    \caption{\footnotesize 
    Two examples of dynamical formation and evolutionary history for merging BBHs involving BH--star collisions. Symbols for the different types of dynamical interactions, including scatterings and exchanges are shown in the inset. Whenever there is a physical collision, we also give the collision time $t_{\rm coll}$. Left side: a BBH undergoes a collision with an MS star (shown in orange) at $t= 16\,$Myr, resulting in the formation of a common envelope. Right side: a massive giant star (shown in red) resulting from previous stellar mergers collides with a single BH at $t= 10\,$Myr; the BH, now spun up by accretion, becomes part of a merging BBH through subsequent dynamical encounters.  }
    \label{fig:cartoon}
\end{figure}

 We show in the left panel the lower limits on BH spins. In this case, only a few BHs can achieve spins up to $\chi \approx 0.4$ through collisions with the most- massive stars while the majority of BHs exhibit low spins ($\chi \lesssim 0.1$). The right panel demonstrates BH spins for the same set of BH--star collisions assuming no wind mass loss ($f_{\rm acc} = 100\%$). In this most optimistic scenario, the median BH spin is 0.4 with approximately a dozen BHs per cluster reaching maximal spins ($\chi= 1$). These results show that stellar collisions in star clusters could significantly enhance the spins of BHs, with important implications for the GW sources detected by LVK, which we will discuss in the next Section.

\section{Implications for Gravitational wave Sources}
\label{sec:gw} 
In this section, we examine the fraction of BHs in merging binaries that are influenced by stellar collisions and, consequently, may experience spin-up.
Figure \ref{fig:cartoon} illustrates the dynamical evolutionary history of two example BBHs, from their birth to the onset of mergers. This figure highlights two types of collision configurations involving BHs and stars, along with the expected spins and orientations of the BHs in each case. On the left side (Case A), a BBH collides with a MS star of radius ($\approx 10\,R_{\odot}$) comparable to the semi-major axis of the orbit ($a\approx 0.1\,$AU), resulting in the BBH becoming embedded in a ``common envelope''-like scenario. In this case, both BHs are expected to accrete stellar debris, which can align their spins with their orbit. If the BBH system does not widen significantly after the interaction, it may coalesce before any subsequent encounters can affect the spin-orbit orientation. On the right side (Case B), a single BH collides with a massive giant which formed from three previous stellar collisions between massive stars. After spinning up through accretion, the BH subsequently forms a BBH through a dynamical exchange encounter at $t=275$~Myr and then hardens via scattering interactions until the merger. For Case B, since the BBH is assembled after the collision that leads to spin up, we expect random orientations of the spins of the BHs at merger.

Table \ref{table:mergers} presents data on BBH mergers affected by prior stellar collisions (Columns 4 and 5), comparing these with those that did not experience any collisions beforehand (Column 3). Specifically, Column 4 shows the number of BBH mergers similar to Case B in Figure~\ref{fig:cartoon}. In this scenario, at least one of the components was spun up through a prior stellar collision as a single BH, after which it forms the BBH through subsequent dynamical encounters, which randomizes the spin-orbit alignment prior to the ultimate BBH merger. Column 5 shows the number BBH mergers similar to Case A. Here the BBH undergoes a stellar collision and merges before undergoing any subsequent dynamical encounters, allowing spin-orbit alignment created during the collision itself to be preserved. In practice, this often means the Case A BBHs are relatively compact at the time of stellar collision (typically $a \lesssim R_{\star}$).
Column 6 provides the total count of BBH mergers, including both those occurring within their host clusters and those that merged post-ejection, represented as the sum of the values in Columns 3 through 5.

In all models combined, roughly $40\%$ of all BBH mergers feature at least one stellar collision prior to merger. About $90\%$ of these spun up BBHs are dynamically assembled from single BHs post-collision, suggesting that the majority of these BBHs have uncorrelated spins and orbital orientations (Case B). In contrast, the remaining $10\%$ experienced a stellar collision after binary formation, thus preserving their spin-orbit alignment until the merger (Case A).

Columns 7-9 represent the number of merging BH binaries with primary, secondary, or both components spinning (due to both BH mergers and stellar collisions) prior to the merger. Given that typical 2G mergers constitute about $10\%$ of all BBH mergers, stellar collisions can increase the number of BBHs with at least one component spinning by a factor of a few (depending on the accretion efficiency; see Figure~\ref{fig:spin_contour}). Additionally, we find that these clusters can host a comparable number of merging BBHs where either the primary component or both components are spinning as a result of prior stellar collisions.

\begin{figure}
    \centering
    \includegraphics[width=0.9\linewidth]{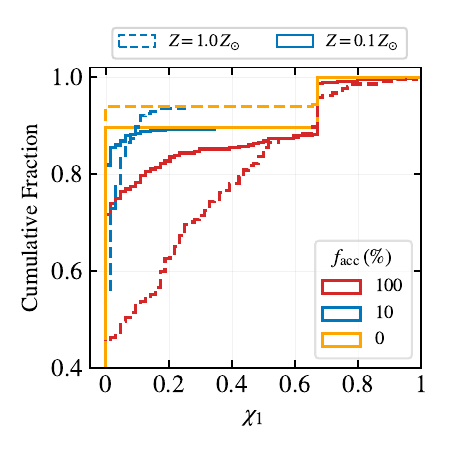}
    \caption{\footnotesize The cumulative spin magnitude distribution of primary BHs in merging binaries, combining all models with $f_{\rm b}=100\%$ (and showing separately results for models with solar- metallicity as dashed lines and lower-metallicity models with solid lines). Different colors indicate three different choices of accretion efficiency (for $f_{\rm acc}=0\%$ BH spin up occurs only through BBH mergers). 
        \label{fig:spin_dist}
    }
\end{figure}

\begin{figure}[h]
    \centering
    \includegraphics[width=1\linewidth]{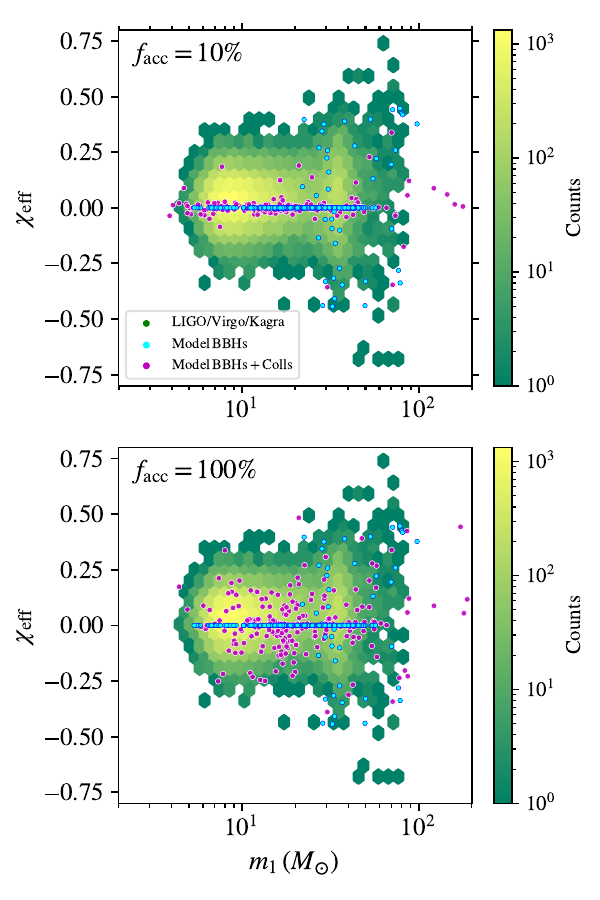}
    \caption{\footnotesize Effective spin ($\chi_{\rm eff}$) vs. primary BH mass for all the merging BBHs identified in our cluster models with $f_{\rm b}(>15\, M_{\odot})=100\%$. Magenta points represent BBHs affected by stellar collisions (with at least one spinning component prior to merger). Blue points denote merging BBHs where no component was affected by a collision (but note that some include a 2G BH). The magenta points differ between the top ($f_{\rm acc}=10\%$) and bottom ($f_{\rm acc}=100\%$) panels. The background color coding shows the 2D posterior predictive distribution in primary mass $m_1$ (source) and $\chi_{\rm eff}$ from LVK data. These were obtained under a population model that allows for the mean and width of the  $\chi_{\rm eff}$  distribution modeled as a Gaussian to correlate linearly with primary mass \citep[Eq.~2--4 of][based on the GWTC-3 data]{Biscoveanu_2022}. }
        \label{fig:chi_eff}
\end{figure}

In Figure \ref{fig:spin_dist}, we show the corresponding cumulative distribution of primary spins of the merging BBHs identified in our high-metallicity (dashed lines) and low-metallicity models (solid lines). Different colors denote different accretion efficiencies (see Section~\ref{sec:spin_calc}). In the case of no accretion (shown in orange), the spin distribution, as expected, peaks at 0 and 0.7, representing 1G and 2G BHs, respectively. As we increase the accretion efficiency, the spin distribution broadens to both lower and larger values than 0.7. For the $Z=1\,Z_{\odot}$ models, where around $60\%$ of all merging BBHs have undergone a stellar collision, the $100\%$ accretion efficiency results in a mean primary spin of approximately $\chi_1\approx 0.2$, while in the $Z=0.1\,Z_{\odot}$ models, the mean primary spin is about $\chi_1\approx 0.1$, with $30\%$ of all merging BBHs having experienced a collision. At a lower accretion efficiency of $10\%$, the mean spin decreases to approximately $\chi_1 \sim 0.01$ for both metallicity models.

In Figure \ref{fig:chi_eff}, we display the effective spin versus primary mass for all of the BBH mergers identified in our models with $f_{\rm b}(>15\, M_{\odot})=100\%$, assuming either 
$10\%$ (top) or $100\%$ (bottom) accretion efficiency. Magenta points indicate BBHs with at least one component that experienced at least one stellar collision before merging, while blue points represent 1G and 2G BBH mergers unaffected by stellar collisions. We compare our results with astrophysical samples from the 2D posterior predictive distribution (PPD) inferred using data from the latest LVK catalog under a population model that allows the $\chi_{\mathrm{eff}}$ distribution to evolve linearly with primary mass. The PPD represents the updated underlying population-level distribution corrected for GW detection biases marginalized over statistical uncertainties \citep[see Appendix D of][]{Thrane_2020}.

We find that our models including the effects of stellar collisions qualitatively reproduce the trends in the GW data. While the $f_{\mathrm{acc}}=10\%$ model predicts a narrower $\chi_{\mathrm{eff}}$ distribution at lower masses, the $f_{\mathrm{acc}}=100\%$ model reproduces both the spread in the $\chi_{\mathrm{eff}}$ values and the broadening of the distribution as a function of mass. The widening at larger masses (blue points) is also due to the merger products of higher-generation BHs.
Preliminary evidence for this broadening was identified in GWTC-3 data, namely, that the width of the $\chi_{\mathrm{eff}}$ distribution increases with the BBH primary mass \citep{Safarzadeh_2020,Biscoveanu_2022,Tiwari_2022,Heinzel_2024}. We caution that the astrophysical population of BBHs probed by LVK detections likely includes contributions from multiple formation channels, so our models should not be expected to explain all of the features in the inferred BBH spin distribution, like the preference for a peak above $\chi_{\mathrm{eff}}=0$. 

 \section{Conclusions and Discussion}
 \label{sec:conc}
  
 \subsection{Summary}
 
BBHs dynamically formed in dense star clusters are expected to evolve through a series of interactions, including collisions and mergers with stars and other BHs. These interactions can result in changes to the BH mass, spin, and orbital characteristics within the cluster. To investigate these effects, we conducted a series of $N$-body simulations of dense star clusters with varying metallicities, densities, and primordial binary fractions. Our primary findings are as follows:

 \begin{enumerate}

\item 
The dynamical evolution of dense star clusters, influenced by their varying initial properties, significantly affects the rates and types of stellar interactions. Depending on these cluster characteristics, collisions between BHs and massive stars can occur both at early and late times, driven by binary evolution and stellar interactions.

 \item The high density in cluster cores combined with high binary fractions lead to an increased rate in stellar mergers and BH--star collisions as binaries have a larger cross section for interaction. Therefore, BHs collide with more-massive stars as we increase the high-mass binary fraction, leading to a higher formation efficiency of spinning BHs (since we assume BHs to be born nonspinning). 

\item We investigated the effect of metallicity on both the number of collisions and BBH mergers. We find that fewer BHs are retained at formation for higher metallicity (due to effects of natal kicks), leading to a reduced BBH merger rate in these simulations. However, we find that the BH--star collision rate \textit{increases} at higher metallicities. This is primarily because lower-mass BHs and stars can mix more readily in the absence of BH-dominated dynamics, particularly as the cluster approaches core collapse. Consequently, despite their lower overall number, more than half of the BHs in our solar-metallicity models can undergo collisions with stars, surpassing the rates obtained in lower-metallicity ($Z=0.1\,Z_{\odot}$) models by at least $50\%$. However, lower-metallicity clusters enable BHs to collide with much more massive stars at early times, allowing for more significant spin-up of the BHs.

\item We explored the formation of spinning BHs through accretion following the collisions and tidal disruptions of stars. We found that initially nonrotating BHs can be spun up to a median dimensionless spin parameter $\chi$ up to about 0.4 in clusters with binary fraction for massive stars $f_{b}(>15\,M_{\odot})=100\%$ (see the right panel of Fig. \ref{fig:spin_contour}, where  $f_{\rm acc} = 100\%$). We also showed that over $50\% $ of merging BBHs in star clusters contain at least one component that has previously collided with a star, resulting in the primary component's median spin $\chi_1$ being up to 0.2, especially in high-metallicity clusters. Formation of highly spinning BHs in young star clusters has key implications for the GW sources detected by the LVK. Indeed, \cite{Biscoveanu_2022} already found some evidence in GW data for a population of BHs with larger spin magnitudes at higher redshift, consistent with our theoretical expectations for dynamically formed BBH sources interacting with massive stars.

 \end{enumerate}

\subsection{Discussion and Future Work}

Close encounters between BHs and stars can also significantly impact BH dynamics within clusters. Merging binary BBHs with high spins can receive GW recoil kicks as large as 4000 km/s \citep[e.g.,][]{Campanelli_2007,Lousto_2012}, which far exceed the escape velocity of any star cluster. Consequently, if many BHs in a cluster achieve high spins through collisions with massive stars, BBH merger products may be more likely to be ejected from the cluster, thereby limiting 2G mergers. \cite{Rodriguez_2019} demonstrated that the fraction of BBH merger products retained within the cluster decreases from approximately $60\%$ to $10\%$ as the birth spins of BHs increase from 0 to 0.2. Similarly, \cite{Antonini_2016} showed that for BH spins bigger than 0.5, 2G mergers would only be formed in massive galactic nuclei. Our $N$-body simulations show that up to $40\%$ of merging BBH components can acquire spins greater than 0.2 through accretion (depending on the assumed accretion efficiency), thereby surpassing the retention of their subsequent mergers.

    BH--star collisions also provide a possible way to produce massive mergers like GW190521 and very asymmetric mergers like GW190412. In the context of these, \cite{Rodriguez_2020A} explored the possibility that GW190412 with the dimensionless spin magnitude of the primary BH $0.43 \pm_{-0.26}^{+0.16}$ \citep{Abbott_2020} is a 3G BBH from a super-star cluster, demonstrating that the primary BH can have a spin magnitude between 0.17 and 0.59. This phenomenon, however, limits the formation of moderately spinning BHs to the retention of 3G BHs in massive clusters ($\gtrsim 10^6\,M_{\odot}$) characterized by large escape speeds. As shown in this paper, BHs with similar spins may be easily produced by invoking stellar collisions in clusters with a typical Milky Way globular cluster mass ($\sim 10^5\,M_{\odot}$).

One interesting question raised by this work is whether the $\chi_{\rm eff}$ distribution could be shifted to slightly positive values through BBH--star collisions in star clusters. \cite{Lopez_2019} investigated this using hydrodynamic simulations and demonstrated how both the spin magnitude and orientation of the BHs can be altered through the accretion of stellar debris, but they did not provide conclusive results regarding any tendency for alignment. A follow-up exploration of this question will be presented in a forthcoming paper (K{\i}ro\u{g}lu et al. 2025, in preparation), which will include detailed hydrodynamic calculations.

Collisions of BBHs with stars can also alter the properties of the BBH orbit, which could potentially affect their GW inspiral time. Collisions of BBHs with giants or massive MS stars may increase their merger rates by shrinking the BBH orbit inside a common envelope. Capturing the detailed tidal interactions during these close encounters requires detailed hydrodynamic calculations, which are beyond the scope of this study (but see, e.g., \cite{Lopez_2019,Kremer_2022,Ryu_2022}, K{\i}ro\u{g}lu et al. 2025, in preparation, for a discussion).

In our cluster models, we have not included primordial mass segregation, although it is observed in many young massive clusters with ages much less than their relaxation times, suggesting it might be a primordial feature of some clusters \citep[e.g.,][]{Bonnell1998,Gouliermis2004,Goswami2012}. Recent studies have shown that primordial mass segregation leads to an increased rate of massive star collisions at early times, thereby facilitating the formation of intermediate-mass BHs through collisional runaway \citep[e.g.,][]{Gurkan_2004,PortegiesZwart_2002,Freitag2006,Goswami2012,Kremer_2020,Gonzalez_2021,Gonzalez_2024,Sharma_2024}. We anticipate that primordial mass segregation could also enhance the rate of BH--star collisions, similar to the effect of primordial binaries.

As this paper represents an initial effort to study BH spin evolution through accretion in realistic stellar environments, several simplifying assumptions have been made. For simplicity, we assume that stars are initially on parabolic orbits and are fully disrupted after any collision or tidal disruption, with $50\%$ of their mass somewhat arbitrarily assumed to remain bound to the BH. The actual amount of stellar material left in an envelope bound to the BH, however, will of course depend on the specific parameters of the encounter. Additionally, we assume that only one of the BHs is able to accrete following a BBH--star collision, while in reality both BHs may be able to accrete some of the debris.   

Another simplifying assumption we make implicitly here is to neglect accretion feedback. Accretion will release energy on a variety of timescales and possibly trigger various outflows, leading to some fraction of mass loss, which is uncertain. In the highly simplified models presented here, we adopt fixed accretion efficiencies of $10\%$ versus $100\%$ to crudely represent accretion under high or low feedback effects.

All of the simplifying assumptions discussed above can also impact the star cluster models that we compute with \texttt{CMC}. 
The current treatment of physical collisions and tidal disruptions in \texttt{CMC} models employs a highly simplified set of recipes. For example, \texttt{CMC} assumes that stars, regardless of their mass, are instantaneously destroyed upon colliding with a BH, with no mass being accreted. Changing BH masses and spins through accretion will actually affect to some extend the long-term dynamical evolution and present-day structure of a star cluster since we now understand that BHs play a dominant role in supporting clusters against gravothermal collapse \citep[e.g.,][]{Kremer_2018_ngc3201,Kremer2020_bhburning}.  

A more detailed treatment of the various regimes of BH–star interactions using hydrodynamic models is necessary to explore the potentially broad range of outcomes from these events. Future work would replace the current simplistic recipes in \texttt{CMC} with a more sophisticated set of fitting formulae, providing at least qualitatively accurate results for all hydrodynamic stellar interactions occurring in clusters.

\begin{acknowledgements}
We thank Selma E. de Mink, Eric Thrane,  Vishal Baibhav, Nick Kaaz, and Zoheyr Doctor for useful discussions.
This work was supported by NSF grant AST-2108624 and NASA grant 80NSSC22K0722, as well as the computational resources and staff contributions provided for the \texttt{Quest} high-performance computing facility at Northwestern University. \texttt{Quest} is jointly supported by the Office of the Provost, the Office for Research, and Northwestern University Information Technology.  F.K. acknowledges support from a CIERA Board of Visitors Graduate Fellowship. S.B. is supported by NASA through the NASA Hubble Fellowship grant HST-HF2-51524.001-A awarded by the Space Telescope Science Institute, which is operated by the Association of Universities for Research in Astronomy, Inc., for NASA, under contract NAS5-26555. Support for E.G.P.\ was provided by the National Science Foundation Graduate Research Fellowship Program under grant DGE-2234667

\end{acknowledgements}

\bibliographystyle{aasjournal}
\bibliography{mybib}

\end{document}